# Quantization of electromagnetic modes and angular momentum on plasmonic nanowires

**Guodong Zhu (朱国栋) [a], Yangzhe Guo (郭杨喆) [a], Bin Dong (董斌) [b,*] and Yurui Fang (方蔚瑞) [a, *]**

*[a.] Key Laboratory of Materials Modification by Laser, Electron, and Ion Beams (Ministry of Education); School of Physics, Dalian University of Technology, Dalian 116024, P.R. China.*

*[b.] Key Laboratory of New Energy and Rare Earth Resource Utilization of State Ethnic Affairs Commission, Key Laboratory of Photosensitive Materials & Devices of Liaoning Province, School of Physics and Materials Engineering, Dalian Nationalities University, 18 Liaohe West Road, Dalian, 116600, PR China*

** Corresponding emails: dong@dlnu.edu.cn (B. Dong), yrfang@dlut.edu.cn (Y. Fang)*

**Abstract:** Quantum theory of surface plasmons is very important for studying the interactions between light and different metal nanostructures in nanoplasmonics. In this work, using the canonical quantization method, the SPPs on nanowires and their orbital and spin angular momentum are investigated. The results show that the SPPs on nanowire carry both orbital and spin momentum during propagation. Later, the result is applied on the plasmonic nanowire waveguide to show the agreement of the theory. The study is helpful for the nano wire based plasmonic interactions and the quantum information based optical circuit in the future.

## 1. Introduction

It is well known that surface plasmons polaritons (SPPs) existing at interfaces between metals and dielectrics are coherent collective oscillation of free electrons (with coupled electromagnetic field) at the surface of metal. The unique properties of plasmons on nanoscale metallic systems have produced a number of dramatic effects and interesting applications, such as molecule detection with surface-enhanced Raman scattering[1-3], biosensing[4, 5], waveguiding[6-8], enhanced interactions[9, 10] and switching devices below the diffraction limit[11, 12]. Plasmonic nanowires have been attracting a lot of attention as SPP waveguides, analogy to optical fiber waveguides but within a hundred nanometers scale cross section, which breaks diffraction limit. The strong confinement and small mode volume of plasmonic wires facilitate the strong coupling between quantum emitters and nanowire[13, 14], the low Q factor Fabry-Pérot resonator[15] and nanowire-wire based plasmonic devices and chips[11, 16-18]. The properties of keeping quantum information like entanglement of plasmonic waveguides[19-22] now have the potential applications in quantum information and fundamental researches.

Surface plasmons are quantized electric charge density waves, while usually a lot of the phenomenon can be illustrated with classical Maxwell equations in this boosting area in the past twenty years[23]. In the meantime, quantum theory descriptions were also developed to explain the phenomenon such as non-locality[24], tunneling[25, 26], hot electrons[27-29] and so on[30]. Especially quantization of surface plasmons for common systems like plane metal interface[31] and nanoparticle[32] has been developed to deal with the interactions between the plasmons and the surrounding molecules[33-35], which is, in a lot of conditions, beyond the Maxwell's theory. Now quantum plasmonics has been a rapid growing field that involves the study of the quantum properties of SPs and its interaction with mater at the nanoscale[36].

The quantization process was usually performed by quantizing the electric field or considering the hydrodynamics meantime. Using Hopfield's approach[37], Elson and Ritchie reported the quantization scheme for SPPs on a metallic surface considering the hydrodynamics[38]. Archambault et al. reported the quantization scheme of surface wave on a plane interface without any specific model of the dielectric constant[39]. Huttner and Barnett introduced a new quantization method by extending Hopfield's approach to polaritons in dispersive and lossy media[40]. In Archambault's works, they redid the quantization of the plasmons on plane interface in modern fashion for explaining the spontaneous and stimulated emission of SPPs[39]. Waks have re-quantized plasmons on nanoparticles under quasi-static approximation for describing the coupling of a dipole and a particle. And in Boardman's work the SPPs in different typical coordinates in common way with Bloch hydrodynamical model[41]. However, quantization of SPPs on cylindrical nanowires needs more investigation to illustrate more details and phenomenon.

In this paper, a quantization scheme of SPPs on a cylindrical nanowire waveguide is introduced. The canonical quantization method is used, which is to identify the generalized coordinates and conjugate momenta. After getting the expressions in Fock states, the orbital and spin angular momentum are deduced. With the results one will find that the modes on the nanowire waveguide carry both orbital and spin angular momentum which is consistent with the classic theory result from Picardi[42]. Following that the formula are applied to describe the SPPs on nanowire in the waveguide which shows good agreement. With the properties, SPPs on nanowires can easily carry quantum information and keep the entangle properties of incident light [21, 43-45]. The results will be very helpful in the quantum information research and light-matter interactions.

## 2. Quantization description of surface waves

Let us consider SPPs propagating on the interface of cylindrical metal nanowire ($\varepsilon_2 = 1 - \frac{\omega_p^2}{\omega^2}$, the imaginary part is ignored because in the resonance range, the dissipation is small, which is also for avoiding the non-orthogonality of the expending bases.) along z axis with radius R in homogeneous lossless dielectric medium ($\varepsilon_1 > 0$). The electric and magnetic fields satisfy

$$\nabla^2 \begin{Bmatrix} \boldsymbol{E} \\ \boldsymbol{H} \end{Bmatrix} - \mu\varepsilon \frac{\partial^2}{\partial t^2} \begin{Bmatrix} \boldsymbol{E} \\ \boldsymbol{H} \end{Bmatrix} = 0 \tag{1}$$

The solutions of $\boldsymbol{E}$ and $\boldsymbol{H}$ in cylindrical coordinates can be deduced in standard steps (Appendix A)[14].

To simplify the process in the following, the electromagnetic scalar and vector potentials $\emptyset$ and $\mathbf{A}$ working in Coulomb gauge ($\emptyset = 0, \nabla \cdot \mathbf{A} = 0$) are introduced here. With $\mathbf{E}(\mathbf{r}, \mathrm{t}) = -\frac{1}{c}\frac{\partial \boldsymbol{A}(\boldsymbol{r},t)}{\partial t}$ and $\mathbf{H}(\mathbf{r}, \mathrm{t}) = \frac{1}{\mu_0} \boldsymbol{\nabla} \times \mathbf{A}(\mathbf{r}, \mathrm{t})$ , the vector potential $\mathbf{A}(\mathbf{r}, \mathrm{t})$ expansion over modes in a volume $V$ can be expressed as

$$\boldsymbol{A}(\boldsymbol{r}, t) = \frac{1}{\sqrt{V}} \sum_{k,m} \boldsymbol{A}_{k,m}(t) e^{i\boldsymbol{k}\cdot\boldsymbol{r}} \tag{2}$$

$$\boldsymbol{A}_{k,m}(t) e^{i\boldsymbol{k}\cdot\boldsymbol{r}} =$$

$$-\frac{i}{\omega} \left\{ \begin{matrix} \left[ \frac{im}{k_{j,m}\rho} a_{j,m} F_{j,m}(k_{j,m\perp}\rho) + \frac{ik_{\parallel,m} k_{j,m\perp}}{k_{j,m}^2} b_{j,m} F'_{j,m}(k_{j,m\perp}\rho) \right] \hat{\rho} \\ + \\ + \left[ -\frac{k_{j\perp}}{k_j} a_{j,m} F'_{j,m}(k_{j,m\perp}\rho) - \frac{mk_{\parallel}}{k_j^2 \rho} b_{j,m} F_{j,m}(k_{j,m\perp}\rho) \right] \widehat{\emptyset} \\ + \frac{k_{j,m\perp}^2}{k_{j,m}^2} b_{j,m} F_{j,m}(k_{j,m\perp}\rho) \hat{z} \end{matrix} \right\} e^{i(m\emptyset + k_{\parallel,m} z)} e^{-i\omega t}$$

$$= \boldsymbol{A}_{k,m}(t) e^{i\boldsymbol{k}_{\parallel,m}\cdot\boldsymbol{r}} = \boldsymbol{A}_{k,m}(t) e^{i\boldsymbol{k}_m\cdot\boldsymbol{r}} \tag{3}$$

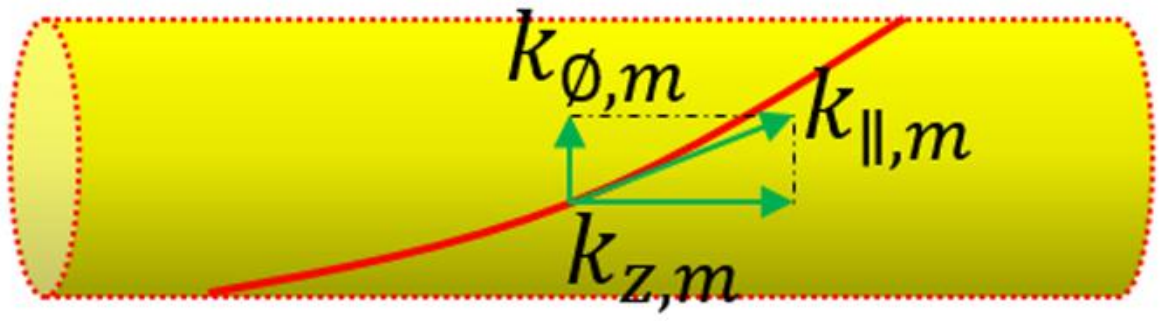

**Fig. 1.** Schematic illustration of SPPs propagate on plasmonic cylindrical waveguide, $\boldsymbol{k}_m \equiv \boldsymbol{k}_{\parallel,m} = 0 \cdot \hat{\rho} + mk_{\emptyset}\widehat{\emptyset} + k_{z,m}\hat{z}$.

where $k_{j,m\perp} = \sqrt{k_j^2 - k_{\parallel,m}^2}$, $\boldsymbol{k}_m \equiv \boldsymbol{k}_{\parallel,m}$ is defined just for simple**,** the phase factor $e^{i\boldsymbol{k}_m\cdot\boldsymbol{r}} = e^{i(m\emptyset + k_{\parallel,m} z)}$ shows that the surface waves have both components on $z$ and $\emptyset$ directions (Fig. 1). $\boldsymbol{k}_m = 0 \cdot \hat{\rho} + mk_{\emptyset}\widehat{\emptyset} + k_{z,m}\hat{z}$, $\boldsymbol{r} = \rho\hat{\rho} + \emptyset\rho\widehat{\emptyset} + z\hat{z}$. Here $k_{\emptyset} = 1/\rho$ and $r_{\emptyset} = \emptyset\rho$ are set to keep the dimensional consistency The subscript $j =$ $1, 2$ represents the region outside and inside of the cylinder nanowire boundary characterized by different dielectric function. $m$ is the azimuthal quantum number. The functions $F_{1,m}(x) = H_m(x)$ and $F_{2,m}(x) = J_m(x)$ are Hankel and Bessel functions. $a_{j,m}$ and $b_{j,m}$ are arbitrary coefficients, which can be fixed by imposing boundary conditions of interface between the wire and surrounding dielectric. $\boldsymbol{A}_{k,m}(t)$ is time-dependent amplitude and can be written as $\mathbf{A}_{\boldsymbol{k},\boldsymbol{m}}\, e^{-i\omega t}$. Because each mode in the waveguide can be represent by order m, in the following we use $m$ to replace $k_m$ and $\boldsymbol{A}_m$ to replace $\mathbf{A}_{\boldsymbol{k},\boldsymbol{m}}$ for the modes subscript. Due to the normalization condition

$$\int d^3r\, e^{i\boldsymbol{k}_m\cdot\boldsymbol{r}} e^{-i\boldsymbol{k}_{m'}\cdot\boldsymbol{r}} = V\delta_{\boldsymbol{mm'}} \tag{4}$$

and the orthogonal of Bessel function, different modes are orthogonal. The amplitudes $\mathbf{A}_{\boldsymbol{m}}$ are complex. Complex conjugate item is added to make the field be real:

$$\boldsymbol{A}(\boldsymbol{r},t) = \frac{1}{2\sqrt{V}} \sum_{k,m} \left[ \boldsymbol{A}_m(t) e^{i\boldsymbol{k}_m\cdot\boldsymbol{r}} + c.c. \right] \tag{5}$$

The canonical momentum $\boldsymbol{p}(\boldsymbol{r},t) = \frac{1}{4\pi c^2}\dot{\boldsymbol{A}}$ satisfies the commutation relation [46]

$$[\boldsymbol{A}_m(\mathbf{r},\mathrm{t}), \boldsymbol{p}_{m'}(\boldsymbol{r}',t)] = i\hbar\delta_{m,m'}\delta(\boldsymbol{r}-\boldsymbol{r}') \tag{6}$$

Now define the cannonical coordinates and momentum as

$$\boldsymbol{Q}_m = \sqrt{\frac{1}{4\pi c^2}}\ (\ \boldsymbol{A}_m + \boldsymbol{A}_m{}^*\ ) \tag{7}$$

$$\boldsymbol{P}_m = -\frac{\mathrm{i}\omega}{\sqrt{4\pi c^2}}\ (\ \boldsymbol{A}_m - \boldsymbol{A}_m{}^*\ ) \tag{8}$$

which makes

$$\boldsymbol{A}(\boldsymbol{r},t) = \frac{\sqrt{4\pi c^2}}{2\sqrt{V}} \sum_m [\boldsymbol{Q}_m cos\boldsymbol{k}\cdot\boldsymbol{r} - \frac{1}{\omega_m}\boldsymbol{P}_m sin\boldsymbol{k}\cdot\boldsymbol{r}] \tag{9}$$

$$\boldsymbol{E}(\boldsymbol{r},t) = -\frac{1}{c}\frac{\partial \boldsymbol{A}(\boldsymbol{r},t)}{\partial t} = -\frac{\sqrt{4\pi}}{2\sqrt{V}} \sum_m \omega_m [\boldsymbol{Q}_m sin\boldsymbol{k}\cdot\boldsymbol{r} + \frac{1}{\omega_m}\boldsymbol{P}_m cos\boldsymbol{k}\cdot\boldsymbol{r}] \tag{10}$$

$$\boldsymbol{B}(\boldsymbol{r},t) = \boldsymbol{\nabla}\times\mathbf{A}(\mathbf{r},\mathrm{t}) = -\frac{\sqrt{4\pi c^2}}{2\sqrt{V}} \sum_m \boldsymbol{k}_m \times [\boldsymbol{Q}_m sin\boldsymbol{k}\cdot\boldsymbol{r} + \frac{1}{\omega_m}\boldsymbol{P}_m cos\boldsymbol{k}\cdot\boldsymbol{r}] \tag{11}$$

Integrating the square of electric and magnetic field over the volume V and using orthogonality relation, the total Hamiltonican is

$$\mathrm{H} = \frac{1}{8\pi}\int d^3r\, (|\boldsymbol{E}|^2 + |\boldsymbol{B}|^2) = \frac{1}{2}\sum_m (\boldsymbol{P}_m^2 + \boldsymbol{\omega}_m^2 \boldsymbol{Q}_m^2) \tag{12}$$

Take the commutators for the canonical coordinates and canonical momentum (which can be also got from equ. (6)) as

$$[\boldsymbol{Q}_m, \boldsymbol{P}_m] = i\hbar \tag{13}$$

Now introduce the creation and annihilation operators

$$\boldsymbol{a}_m = \frac{1}{\sqrt{2\hbar\omega}}(\omega_m \boldsymbol{Q}_m + i\boldsymbol{P}_m) \tag{14}$$

$$\boldsymbol{a}_m^\dagger = \frac{1}{\sqrt{2\hbar\omega}}(\omega_m \boldsymbol{Q}_m - i\boldsymbol{P}_m) \tag{15}$$

Based on the definitions, we have

$$[\boldsymbol{a}_m, \boldsymbol{a}_m^\dagger] = \frac{1}{2\hbar\omega_m}[\omega_m \boldsymbol{Q}_m + i\boldsymbol{P}_m, \omega_m \boldsymbol{Q}_m - i\boldsymbol{P}_m]$$

$$= \frac{1}{2\hbar\omega_m}(-i\omega_m[\boldsymbol{Q}_m, \boldsymbol{P}_m] + i\omega_m[\boldsymbol{P}_m, \boldsymbol{Q}_m]) = 1 \quad (16)$$

Considering the positive and negative wave vector sums, the quantized vector potential, electromagnetic fields and Hamiltonian are

$$\boldsymbol{A}(\boldsymbol{r},t) = \sqrt{\frac{2\pi c^2\hbar}{V}}\sum_m \frac{1}{\sqrt{\omega_m}}[\boldsymbol{a}_m e^{i\boldsymbol{k}_m\cdot\boldsymbol{r}} + \boldsymbol{a}_m^\dagger e^{-i\boldsymbol{k}_m\cdot\boldsymbol{r}}] \quad (17)$$

$$\boldsymbol{E}(\boldsymbol{r},t) = i\frac{\sqrt{2\pi\hbar}}{\sqrt{V}}\sum_m \sqrt{\omega_m}[\boldsymbol{a}_m e^{i\boldsymbol{k}_m\cdot\boldsymbol{r}} - \boldsymbol{a}_m^\dagger e^{-i\boldsymbol{k}_m\cdot\boldsymbol{r}}] \quad (18)$$

$$\boldsymbol{B}(\boldsymbol{r},t) = i\sqrt{\frac{2\pi c^2\hbar}{V}}\sum_m \frac{1}{\sqrt{\omega_m}}\boldsymbol{k}_m \times [\boldsymbol{a}_m e^{i\boldsymbol{k}_m\cdot\boldsymbol{r}} - \boldsymbol{a}_m^\dagger e^{-i\boldsymbol{k}_m\cdot\boldsymbol{r}}] \quad (19)$$

$$\mathrm{H} = \sum_m \mathrm{H}_m = \sum_m \hbar\omega_m(\boldsymbol{a}_m^\dagger \boldsymbol{a}_m + 1/2) = \sum_m \hbar\omega_m(n_m + 1/2) \quad (20)$$

with $n_m = \boldsymbol{a}_m^\dagger \boldsymbol{a}_m$ is the number operator. And the evolution obeys

$$i\hbar\,\dot{\boldsymbol{a}}_m = [\boldsymbol{a}_m, \mathrm{H}] \quad (21)$$

$$i\hbar\,\dot{\boldsymbol{a}}_m^\dagger = [\boldsymbol{a}_m^\dagger, \mathrm{H}] \quad (22)$$

## 3. Polarization and Angular momentum

We set $\hat{\epsilon}_\rho = (1,0,0)$, $\hat{\epsilon}_\emptyset = (0,1,0)$, $\hat{\epsilon}_z = (0,0,1)$. Because $\boldsymbol{A}(\boldsymbol{r},t)$ is a vector which can be expressed as $\boldsymbol{A}(\boldsymbol{r},t) = \sum_\alpha \hat{\epsilon}_\alpha \boldsymbol{A}_\alpha(\boldsymbol{r},t)$, the creation and annihilation operators can be expressed correspondingly as $\boldsymbol{a}_m = \sum_\alpha \hat{\epsilon}_\alpha \boldsymbol{a}_{m,\alpha}(\boldsymbol{r},t)$ $\boldsymbol{a}_m^\dagger = \sum_\alpha \hat{\epsilon}_\alpha^\dagger \boldsymbol{a}_{m,\alpha}^\dagger(\boldsymbol{r},t)$ and we have

$$\boldsymbol{A}(\boldsymbol{r},t) = \sqrt{\frac{2\pi c^2\hbar}{V}}\sum_{m,\alpha} \frac{1}{\sqrt{\omega_m}}[\hat{\epsilon}_\alpha \boldsymbol{a}_{m,\alpha} e^{i\boldsymbol{k}_m\cdot\boldsymbol{r}} + \hat{\epsilon}_\alpha^\dagger \boldsymbol{a}_{m,\alpha}^\dagger e^{-i\boldsymbol{k}_m\cdot\boldsymbol{r}}] \quad (23)$$

$$\boldsymbol{E}(\boldsymbol{r},t) = i\sqrt{\frac{2\pi\hbar}{V}}\sum_{m,\alpha} \sqrt{\omega_m}[\hat{\epsilon}_\alpha \boldsymbol{a}_{m,\alpha} e^{i\boldsymbol{k}_m\cdot\boldsymbol{r}} - \hat{\epsilon}_\alpha^\dagger \boldsymbol{a}_{m,\alpha}^\dagger e^{-i\boldsymbol{k}_m\cdot\boldsymbol{r}}] \quad (24)$$

$$\boldsymbol{B}(\boldsymbol{r},t) = i\sqrt{\frac{2\pi\hbar}{V}}\sum_{m,\alpha} \frac{c\boldsymbol{k}_m\times}{\sqrt{\omega_m}}[\hat{\epsilon}_\alpha \boldsymbol{a}_{m,\alpha} e^{i\boldsymbol{k}_m\cdot\boldsymbol{r}} - \hat{\epsilon}_\alpha^\dagger \boldsymbol{a}_{m,\alpha}^\dagger e^{-i\boldsymbol{k}_m\cdot\boldsymbol{r}}] \quad (25)$$

And momentum

$$\boldsymbol{p} = \int d^3r \frac{\boldsymbol{E}\times\boldsymbol{B}}{4\pi c} = \frac{i^2}{4\pi c}\frac{hc}{V}\sum_{m,\alpha}\sum_{m',\alpha'}[\hat{\epsilon}_\alpha \times (\boldsymbol{k}' \times \hat{\epsilon}_{\alpha'})] \times \int d^3r(\boldsymbol{a}_{m,\alpha} e^{i\boldsymbol{k}_m\cdot\boldsymbol{r}} - \boldsymbol{a}_{m,\alpha}^\dagger e^{-i\boldsymbol{k}_m\cdot\boldsymbol{r}})(\boldsymbol{a}_{m',\alpha'} e^{i\boldsymbol{k}_{m'}\cdot\boldsymbol{r}} - \boldsymbol{a}_{m',\alpha'}^\dagger e^{-i\boldsymbol{k}_{m'}\cdot\boldsymbol{r}}) = \frac{h}{4\pi}\sum_{m,\alpha}\boldsymbol{k}_m[\boldsymbol{a}_{m,\alpha}\boldsymbol{a}_{m,\alpha}^\dagger + \boldsymbol{a}_{m,\alpha}^\dagger \boldsymbol{a}_{m,\alpha}] = \sum_m \hbar\boldsymbol{k}_m n_m \quad (26)$$

The zero point term 1/2 is cancelled by the term of $+\boldsymbol{k}_m$ and $-\boldsymbol{k}_m$.

With the above results we will discuss the quantized field properties in the nanowire, such as spin and orbital angular momentum (SAM/OAM) and polarization. It has been known that the optical angular momentum is

$$\boldsymbol{J} = \frac{1}{4\pi c}\int d^3\boldsymbol{r}\,\boldsymbol{r} \times (\boldsymbol{E} \times \boldsymbol{B}) = \frac{1}{4\pi c}\int d^3\boldsymbol{r}[E_\alpha(\boldsymbol{r} \times \boldsymbol{\nabla})A_\alpha + \boldsymbol{E} \times \boldsymbol{A}] \tag{27}$$

The first term of the right hand is the orbit part and the second is the spin part. The orbital angular momentum density on the surface is (Appendix B):

$$\boldsymbol{l_\alpha} = (\boldsymbol{r} \times \boldsymbol{\nabla})A_\alpha|_{\rho=R} = \sum_{\boldsymbol{m}} \hat{r}_\alpha \cdot \begin{pmatrix} -\frac{z}{R}\frac{\partial}{\partial\emptyset}\,\hat{\boldsymbol{\rho}} \\ \left(z\frac{\partial}{\partial\rho} - R\frac{\partial}{\partial z}\right)\hat{\emptyset} \\ \frac{\partial}{\partial\emptyset}\,\hat{\boldsymbol{z}} \end{pmatrix} A_\alpha$$

$$= i\sqrt{\frac{2\pi c^2\hbar}{V}}\sum_m \frac{1}{\sqrt{\omega_m}}\hat{r}_\alpha \cdot R \cdot \begin{pmatrix} -\frac{z}{R}mk_\emptyset \\ -k_{z,m} \\ mk_\emptyset \end{pmatrix}\left[\boldsymbol{a}_{m\alpha}e^{i\boldsymbol{k}_{\parallel,m}\boldsymbol{r}} - \boldsymbol{a}^\dagger_{m\alpha}e^{-i\boldsymbol{k}_{\parallel,m}\boldsymbol{r}}\right]$$

$$= \sqrt{\frac{2\pi c^2\hbar}{V}}\sum\nolimits_{\boldsymbol{m}} \frac{1}{\sqrt{\omega_m}}\,\hat{r}_\alpha \cdot \begin{pmatrix} -\frac{z}{R}m \\ -Rk_{z,m} \\ m \end{pmatrix} \cdot [\boldsymbol{a}_{m\alpha}e^{i\boldsymbol{k}_{\parallel,\boldsymbol{m}}\boldsymbol{r}} - \boldsymbol{a}^\dagger_{m\alpha}e^{-i\boldsymbol{k}_{\parallel,\boldsymbol{m}}\boldsymbol{r}}] \tag{28}$$

With the creation and annihilation operators, the orbital angular momentum is

$$\boldsymbol{L} = \frac{1}{4\pi c}\int \boldsymbol{d^3r}E_\alpha l_\alpha = \frac{i\hbar}{2V}\sum\nolimits_{m,m',\alpha,\alpha'}\frac{\omega_m}{\sqrt{\omega_m\omega_{m'}}}\hat{\boldsymbol{\epsilon}}_{\boldsymbol{\alpha}} \cdot \hat{\boldsymbol{\epsilon}}_{\boldsymbol{\alpha'}}\int d^3\boldsymbol{r}\,R * \begin{pmatrix} -\frac{z}{R}mk_\emptyset \\ -k_{z,m} \\ mk_\emptyset \end{pmatrix}\left[\boldsymbol{a}_{m\alpha}e^{i\boldsymbol{k}_{\parallel,m}\boldsymbol{r}} - \boldsymbol{a}^\dagger_{m\alpha}e^{-i\boldsymbol{k}_{\parallel,m}\boldsymbol{r}}\right]\left[\boldsymbol{a}_{m'\alpha'}e^{i\boldsymbol{k}_{\parallel,m'}\boldsymbol{r}} - \boldsymbol{a}^\dagger_{m'\alpha'}e^{-i\boldsymbol{k}_{\parallel,m'}\boldsymbol{r}}\right]$$

$$= \frac{i\hbar}{2}\sum\nolimits_{m,\alpha}\begin{pmatrix} 0 \\ -Rk_{z,m} \\ m \end{pmatrix}\left(\boldsymbol{a}_{m,\alpha}\boldsymbol{a}^\dagger_{m,\alpha} + \boldsymbol{a}^\dagger_{m,\alpha}\boldsymbol{a}_{m,\alpha}\right) = \sum\nolimits_m i\hbar\boldsymbol{l}_m n_m \tag{29}$$

$$\boldsymbol{l}_m = \begin{pmatrix} 0 \\ -Rk_{z,m} \\ m \end{pmatrix} \tag{30}$$

$$L_z = \frac{\boldsymbol{L}\cdot\boldsymbol{k}_z}{|\boldsymbol{k}_z|} = i\hbar\sum\nolimits_m m * \boldsymbol{n}_m \tag{31}$$

where $n_{m,\alpha} \equiv \boldsymbol{a}^\dagger_{m,\alpha}\boldsymbol{a}_{m,\alpha}$ and $n_m \equiv \boldsymbol{a}^\dagger_m\boldsymbol{a}_m$.

In the above expressions, one can rewrite the phase factor

$$e^{i\boldsymbol{k}_{\parallel,m}\boldsymbol{r}} = e^{i\boldsymbol{k}_m\cdot\boldsymbol{r}} = e^{i(m\emptyset+k_z z)} = [\cos(m\emptyset) + isin(m\emptyset)]e^{ik_z z} \tag{32}$$

$$e^{i\boldsymbol{k}_{\parallel,-m}\boldsymbol{r}} = e^{i\boldsymbol{k}_{-m}\cdot\boldsymbol{r}} = e^{i(-m\emptyset+k_z z)} = [\cos(m\emptyset) - isin(m\emptyset)]e^{ik_z z} \tag{33}$$

which shows that the field rotation is very similar to the circularly polarized light. Then if we absorb the phase factor $e^{im\emptyset}$ into the creation and annihilation operator, we can

directly write

$$\boldsymbol{a}_m e^{i\boldsymbol{k}_m\cdot\boldsymbol{r}} = \sum_\alpha \hat{\epsilon}_\alpha \boldsymbol{a}_{m,\alpha} e^{i\boldsymbol{k}_m\cdot\boldsymbol{r}} = \sum_\alpha \lambda_{+\alpha}\hat{\epsilon}_\alpha \boldsymbol{a}_{m,\alpha} e^{ik_{\parallel,m}\cdot\boldsymbol{r}} = \boldsymbol{a}_{L,m} e^{ik_z z} \quad (34)$$

$$\boldsymbol{a}_{-m} e^{i\boldsymbol{k}_{-m}\cdot\boldsymbol{r}} = \boldsymbol{a}_{R,m} e^{ik_z z} \quad (35)$$

where $L$ and $R$ sign here referring the chirality without further explanation and we will see the physics meaning later. $\boldsymbol{A}_{-m}(t) = -\boldsymbol{A}_m(t)$, $\boldsymbol{A}_m(t)$ (m $= 0, \pm 1, \pm 2, \ldots$) compose over-complete bases and so as $\boldsymbol{a}_m$.

The spin angular momentum density is

$$\boldsymbol{s} = \boldsymbol{E} \times \boldsymbol{A} = \frac{i}{\hbar}(E_\alpha (\boldsymbol{S}_\alpha)_{\beta\gamma} A_\alpha) \quad (36)$$

$$(\boldsymbol{S}_\alpha)_{\beta\gamma} = -\boldsymbol{i}\hbar\boldsymbol{\varepsilon}_{\alpha\beta\gamma} \quad (37)$$

The operator $S_\alpha$ defined here satisfies the commutation relation (Appendix B), which shows that it is a spin operator:

$$[(\boldsymbol{S}_i), (\boldsymbol{S}_j)] = i\hbar\boldsymbol{\varepsilon}_{ijk}(\boldsymbol{S}_k) \quad (38)$$

Calculating $\vec{S}^2$, $\mathrm{s} = 1$ corresponding to spin-1 (SPPs are the quasi-particles) can be obtained:

$$\vec{\boldsymbol{S}}^{\boldsymbol{2}} = (\boldsymbol{S}_i)_{lm}(\boldsymbol{S}_i)_{mn} = (-i\hbar)^{\boldsymbol{2}}(\delta_{ln}\delta_{mn} - \delta_{ln}\delta_{mn}) \quad (39)$$

$$(\vec{\boldsymbol{S}}^{\boldsymbol{2}})_{ln} = (-i\hbar)^{\boldsymbol{2}}(\delta_{ln} - 3\delta_{ln}) = 2\hbar^2\delta_{ln} \quad (40)$$

$$\mathrm{s}(\mathrm{s}+1) = 2, \mathrm{s} = 1 \quad (41)$$

The total spin momentum is

$$\boldsymbol{S} = \frac{1}{4\pi c}\int \boldsymbol{d^3r}\,(\boldsymbol{s}) = \frac{1}{4\pi c}\int \boldsymbol{d^3r}\, E_\alpha(\boldsymbol{S})_{\beta\gamma}A_\alpha \quad (42)$$

$$\boldsymbol{S} = \frac{-1}{4\pi c}\frac{ihc}{V}\sum_{ma}\sum_{m'a'} \varepsilon_{\alpha\beta\gamma}\hat{\boldsymbol{\epsilon}}_\alpha \times \hat{\boldsymbol{\epsilon}}_{\alpha'} \int \boldsymbol{d^3r}\,[\boldsymbol{a}_{m,\alpha}e^{i\boldsymbol{k}_m\cdot\boldsymbol{r}}$$

$$+\boldsymbol{a}^\dagger_{m,\alpha}e^{-i\boldsymbol{k}_{m'}\cdot\boldsymbol{r}}][\boldsymbol{a}_{m',\alpha'}e^{i\boldsymbol{k}_m\cdot\boldsymbol{r}} - \boldsymbol{a}^\dagger_{m',\alpha'}e^{-i\boldsymbol{k}_{m'}\cdot\boldsymbol{r}}] \quad (43)$$

$$\boldsymbol{S} = \frac{ih}{2\pi}\sum_{m\alpha\alpha'} \varepsilon_{\alpha\beta\gamma}\hat{\boldsymbol{\epsilon}}_\alpha \times \hat{\boldsymbol{\epsilon}}_{\alpha'}\,(\boldsymbol{a}_{m,\alpha}\boldsymbol{a}^\dagger_{m,\alpha'} - \boldsymbol{a}^\dagger_{m,\alpha}\boldsymbol{a}_{m,\alpha'}) \quad (44)$$

Because both $m = m$ and $m = -m'$ give the same results, so we put a factor 2 in the last step. Use $\hat{\boldsymbol{\epsilon}}_\alpha \times \hat{\boldsymbol{\epsilon}}_\beta = \hat{\boldsymbol{k}}_\gamma$ $(\alpha, \beta, \gamma = \rho, \emptyset, z)$, we get

$$\boldsymbol{S} = \sum_{m\alpha\alpha'} i\hbar\hat{\boldsymbol{k}}_{m,\alpha''}\,(\boldsymbol{a}_{m,\alpha}\boldsymbol{a}^\dagger_{m,\alpha'} - \boldsymbol{a}^\dagger_{m,\alpha}\boldsymbol{a}_{m,\alpha'}) \quad (45)$$

From the expression one can clearly see that the spin in one direction (like $\emptyset$ direction) is yielded by the other two components ($\rho$ $and$ $z$) of the field. It also agrees with that there is transverse spin ($\boldsymbol{S}_{\perp} = \frac{Re\,\boldsymbol{k} \times Im\,\boldsymbol{k}}{Re\,\boldsymbol{k^2}}$)[47] for surface waves which is represented by the elliptical trace of the field vector at local points.

For plane waves, polarization is usually defined as $\hat{\epsilon}_1 = \hat{x}$ and $\hat{\epsilon}_2 = \hat{y}$. Then circularly polarized light is expressed as $\hat{\boldsymbol{\epsilon}}_L = \hat{\boldsymbol{\epsilon}}_+ = \frac{1}{\sqrt{2}}(\hat{x} + i\hat{y})$ and $\hat{\boldsymbol{\epsilon}}_R = \hat{\boldsymbol{\epsilon}}_- = \frac{1}{\sqrt{2}}(\hat{x} - i\hat{y})$. Following this, here we define[48]

$$\hat{\boldsymbol{s}}_{\alpha,L} = \hat{\boldsymbol{s}}_{\alpha,+} = \frac{1}{\sqrt{2}}\left(\hat{\boldsymbol{\epsilon}}_\beta + i\hat{\boldsymbol{\epsilon}}_\gamma\right) \tag{46}$$

$$\hat{\boldsymbol{s}}_{\alpha,R} = \hat{\boldsymbol{s}}_{\alpha,-} = \frac{1}{\sqrt{2}}\left(\hat{\boldsymbol{\epsilon}}_\beta - i\hat{\boldsymbol{\epsilon}}_\gamma\right) \tag{47}$$

Then

$$\hat{\boldsymbol{\epsilon}}_\beta \boldsymbol{a}_{m,\beta} + \hat{\boldsymbol{\epsilon}}_\gamma \boldsymbol{a}_{m,\gamma} = \frac{1}{\sqrt{2}}\left(\hat{\boldsymbol{s}}_{\alpha,L} + \hat{\boldsymbol{s}}_{\alpha,R}\right)\boldsymbol{a}_{m,\beta} + \frac{1}{i\sqrt{2}}\left(\hat{\boldsymbol{s}}_{\alpha,L} - \hat{\boldsymbol{s}}_{\alpha,R}\right)\boldsymbol{a}_{m,\gamma}$$

$$= \frac{1}{\sqrt{2}}\left(\boldsymbol{a}_{m,\beta} - i\boldsymbol{a}_{m,\gamma}\right)\hat{\boldsymbol{s}}_{\alpha,L} + \frac{1}{\sqrt{2}}\left(\boldsymbol{a}_{m,\beta} + i\boldsymbol{a}_{m,\gamma}\right)\hat{\boldsymbol{s}}_{\alpha,R} \tag{48}$$

So we can define new circularly polarized operators

$$\boldsymbol{a}_{m,L,\alpha} = \frac{1}{\sqrt{2}}\left(\boldsymbol{a}_{m,\beta} - i\boldsymbol{a}_{m,\gamma}\right) \tag{49}$$

$$\boldsymbol{a}_{m,R,\alpha} = \frac{1}{\sqrt{2}}\left(\boldsymbol{a}_{m,\beta} + i\boldsymbol{a}_{m,\gamma}\right) \tag{50}$$

$$\boldsymbol{a}^\dagger_{m,L,\alpha} = \frac{1}{\sqrt{2}}\left(\boldsymbol{a}^\dagger_{m,\beta} - i\boldsymbol{a}^\dagger_{m,\gamma}\right) \tag{51}$$

$$\boldsymbol{a}^\dagger_{m,R,\alpha} = \frac{1}{\sqrt{2}}\left(\boldsymbol{a}^\dagger_{m,\beta} + i\boldsymbol{a}^\dagger_{m,\gamma}\right) \tag{52}$$

With $\left[\boldsymbol{a}_{m,L,\alpha}, \boldsymbol{a}^\dagger_{m,L,\alpha}\right] = 1$ and $\left[\boldsymbol{a}_{m,R,\alpha}, \boldsymbol{a}^\dagger_{m,R,\alpha}\right] = 1$. One can get

$$\boldsymbol{a}_{m,\beta} = \frac{1}{\sqrt{2}}\left(\boldsymbol{a}_{m,L,\alpha} + \boldsymbol{a}_{m,R,\alpha}\right) \tag{53}$$

$$\boldsymbol{a}_{m,\gamma} = \frac{i}{\sqrt{2}}\left(\boldsymbol{a}_{m,L,\alpha} - \boldsymbol{a}_{m,R,\alpha}\right) \tag{54}$$

$$\boldsymbol{a}^\dagger_{m,\beta} = \frac{1}{\sqrt{2}}\left(\boldsymbol{a}^\dagger_{m,L,\alpha} + \boldsymbol{a}^\dagger_{m,R,\alpha}\right) \tag{55}$$

$$\boldsymbol{a}^\dagger_{m,\gamma} = \frac{-i}{\sqrt{2}}\left(\boldsymbol{a}^\dagger_{m,L,\alpha} - \boldsymbol{a}^\dagger_{m,R,\alpha}\right) \tag{56}$$

Obviously

$$\sum\nolimits_{a=\alpha,\beta} \hat{\boldsymbol{\epsilon}}_a \, \boldsymbol{a}_{m,a} = \sum\nolimits_{\gamma}(\hat{\boldsymbol{s}}_{\gamma,L}\boldsymbol{a}_{m,L,\gamma} + \hat{\boldsymbol{s}}_{\gamma,R}\boldsymbol{a}_{m,R,\gamma}) \tag{57}$$

$$\sum\nolimits_{a=\alpha,\beta} \hat{\boldsymbol{\epsilon}}^\dagger_a \boldsymbol{a}^\dagger_{m,a} = \sum\nolimits_{\gamma}(\hat{\boldsymbol{s}}^\dagger_{\gamma,L}\boldsymbol{a}^\dagger_{m,L,\gamma} + \hat{\boldsymbol{s}}^\dagger_{\gamma,R}\boldsymbol{a}^\dagger_{m,R,\gamma}) \tag{58}$$

Now the spin operator becomes

$$\boldsymbol{S} = \sum_{m\alpha} \hbar\widehat{\boldsymbol{k}}_{m,\alpha}(\boldsymbol{a}^{\dagger}_{m,L,\alpha}\boldsymbol{a}_{m,L,\alpha} - \boldsymbol{a}^{\dagger}_{m,R,\alpha}\boldsymbol{a}_{m,R,\alpha}) = \sum_{m\alpha} \hbar\widehat{\boldsymbol{k}}_{m,\alpha}(\boldsymbol{n}_{m,L,\alpha} - \boldsymbol{n}_{m,R,\alpha}) \quad (59)$$

$$\hat{S}_{\alpha} = \frac{\widehat{\boldsymbol{S}}\cdot\boldsymbol{k}_{\alpha}}{|\boldsymbol{k}_{\alpha}|} \quad (60)$$

Then the total angular momentum can be expressed as

$$\boldsymbol{J} = \boldsymbol{L} + \boldsymbol{S} = \sum_{m,\alpha} i\hbar\boldsymbol{k}_{m,\alpha}\widehat{\boldsymbol{n}}_{m,\alpha} + \sum_{m,\alpha} \hbar\widehat{\boldsymbol{k}}_{m,\alpha}(\boldsymbol{n}_{m,L,\alpha} - \boldsymbol{n}_{m,R,\alpha}) \quad (61)$$

## 4. SPPs on silver nanowires as waveguides

The nanowires performed as plasmonic waveguides are usually in homogeneous medium and excited with light perpendicular to the wire axis on one end[6]. The local symmetry at the terminus of wire is broken when the SPPs are excited, so the phase factor $e^{im\emptyset}$ becomes $\cos(m\emptyset)$ or $\sin(m\emptyset)$. For thin wires, the higher-order modes $|m| \geq 2$ are cut off[14]. And the retardation effects are significant, which will cause the excitation phase difference for $|m| = 0$ and $|m| = 1$ modes as shown in Fig. 2[49]. The three lowest modes are excited simultaneously and propagating in a fixed phase delay. When the wire is excited by a beam of light propagating along $-y$ direction with linear polarization in 45° with wire axis, the polarization can be decomposed into $x$ and $z$ directions. The $z$ component will excite $m = 0$ mode and $m' = +1$ mode ($\sin(m\emptyset)$) with $\pi/2$ phase difference. The $x$ component will excite $m = +1$ mode ($\cos(m\emptyset)$) with the same phase of $m = 0$ mode. So the potential vector can be expressed as (only consider the outside of boundary, the inside is similar).

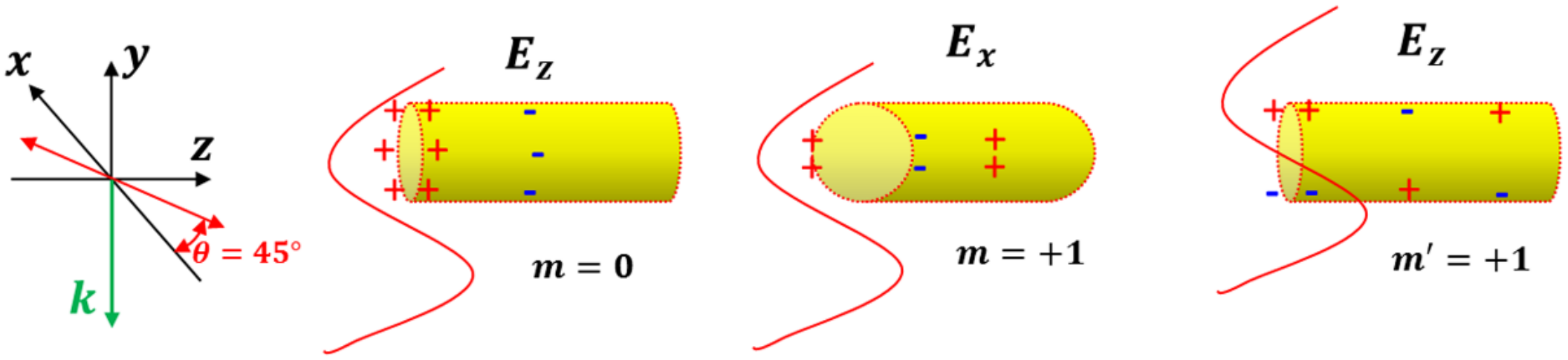

**Fig. 2.** Schematic illustration of the plasmonic cylindrical wire excited by light at one terminal for lower order modes. The light goes along $-y$ direction and the polarization is on 45°with $z$ axis which can be decomposed with $x$ and $z$ components.

$$\boldsymbol{A}_{l=1}(\boldsymbol{r},t) = \boldsymbol{A}_0 e^{i(k_{z,0}z-\omega t)} + \boldsymbol{A}_{+1}\cos(\emptyset)\, e^{i(k_{z,+1}z-\omega t)} + \boldsymbol{A}_{+1}\cos\left(\emptyset - \frac{\pi}{2}\right) e^{i\left(k_{z,+1}z-\omega t+\frac{\pi}{2}\right)} = \boldsymbol{A}_0 e^{i(k_{z,0}z-\omega t)} + \boldsymbol{A}_{+1}\cos(\emptyset)\, e^{i(k_{z,+1}z-\omega t)} + i * \boldsymbol{A}_{+1}\sin(\emptyset)\, e^{i(k_{z,+1}z-\omega t)} = \boldsymbol{A}_0 e^{i(k_{z,0}z-\omega t)} + \boldsymbol{A}_{+1} e^{i\emptyset} e^{i(k_{z,+1}z-\omega t)} \quad (62)$$

It can be rewritten as

$$\boldsymbol{A}_{l=1}^{R}(\boldsymbol{r},t)=\boldsymbol{A}_0 e^{i(k_{z,0}z-\omega t)}+\boldsymbol{A}_{+1}e^{i\emptyset}e^{i(k_{z,+1}z-\omega t)}$$

$$=e^{i(k_{z,0}z-\omega t)}(\boldsymbol{A}_0+\boldsymbol{A}_{+1}e^{i(\Delta kz+\emptyset)}) \quad (63)$$

where $\Delta k=k_{z,+1}-k_{z,0}$, $\Delta k\hat{z}+\emptyset\hat{\emptyset}=\Delta\boldsymbol{k_m}=\boldsymbol{k_{m,+1}}-\boldsymbol{k_{m,0}}$. When the wire is excited by a linear polarized light in $-45°$ can be expressed as

$$\boldsymbol{A}_{l=1}^{L}(\boldsymbol{r},t)=\boldsymbol{A}_0 e^{i(k_{z,0}z-\omega t)}+\boldsymbol{A}_{-1}e^{-i\emptyset}e^{i(k_{z,+1}z-\omega t)}$$

$$=e^{i(k_{z,0}z-\omega t)}(\boldsymbol{A}_0+\boldsymbol{A}_{-1}e^{i(\Delta kz-\emptyset)}) \quad (64)$$

The factor $e^{i(\Delta kz+\emptyset)}$ shows the spiral propagation of the SPPs clearly as reference [49] (shown as Fig. 3), where the coherent interference of SPP waves of $m=+1$ and $m'=+1$ integrate a circularly-polarization-like wave ( $\cos(\emptyset)+i\sin(\emptyset)=e^{i\emptyset}$ ). The superposition with mode $m=0$ will yield beat effect ($e^{i\Delta kz}$), which stretch the circularly-wave into a helical wave as the factor $e^{i(\Delta kz+\emptyset)}$ shown (Fig. 3a). The helical wave is very similar to the vortex wave with orbital angular momentum $l=+1$ (Fig. 3a and 3b $\boldsymbol{A}_1^R$). Analogically, the SPP waves on cylindrical nanowires with higher modes $m=\pm2,\pm3,\dots$ are similar to the vortex wave of $l=\pm2,\pm3,\dots$. The spiral is right handed which was mentioned in the polarization part as $\boldsymbol{a}_{R,m}$. Similarly, the superposition of $m=0$ and $m=-1$ modes will cause a $l=-1$ helical wave (Fig. 3b $\boldsymbol{A}_1^L$). A special case of the condition is when the incident polarization is along the wire. Then

$$\boldsymbol{A}_{l=1}^{0}(\boldsymbol{r},t)=\boldsymbol{A}_0 e^{i(k_{z,0}z-\omega t)}+\boldsymbol{A}_{+1}\sin(\emptyset)\,e^{i(k_{z,+1}z-\omega t)}$$

$$=e^{i(k_{z,0}z-\omega t)}(\boldsymbol{A}_0+\sin(\emptyset)\,\boldsymbol{A}_{+1}e^{i\Delta kz}) \quad (65)$$

which shows only a beat in $\emptyset=\pi/2$ side of the wire (Fig. 3b $\boldsymbol{A}_1^0$).

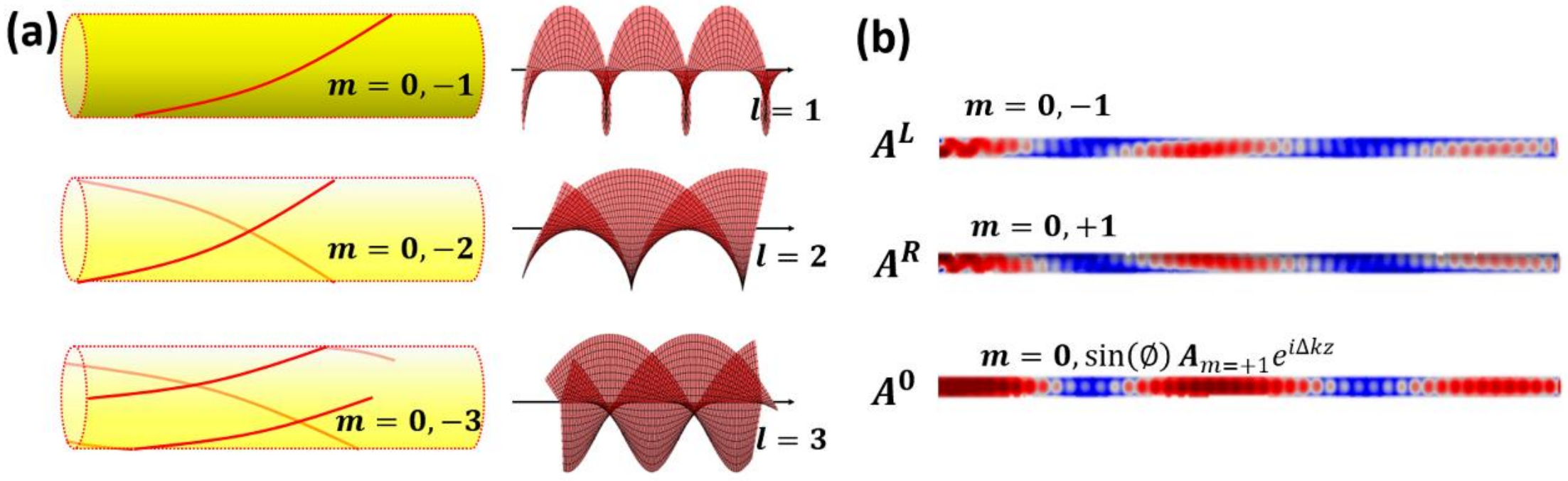

**Fig. 3.** (a) Schematic illustration of the superposed $m = 0$ mode and $m > 0$ modes, which is similar to the vortex waves with $l = 1,2,3$. The red lines represent the maximum of the beats waves; the lighter lines indicate the beats behind the wire and the darker ones indicate the beats on the reader's side. (b) The scheme of electric field distributions on plasmonic cylindrical nanowires excited by focused light one left terminals with polarization in $+45°(\mathrm{A}^L)$, $-45°(\mathrm{A}^R)$ and $0°(\mathrm{A}^0)$ directions to $z$ axis.

With the vector potential one can directly achieve the following results in Fock states.

$$\mathrm{H}^L|n\rangle_m = \sum_{\substack{m=0,-1\\ a}} \hat{\boldsymbol{\epsilon}}_\alpha \hbar\omega_m\left(\boldsymbol{a}_{m,a}^\dagger \boldsymbol{a}_{m,a} + 1/2\right)|n\rangle_m = \sum_{m=0,-1} \hbar\omega_m(n_m + 1/2) \quad (66)$$

$$\mathrm{H}^R|n\rangle_m = \sum_{\substack{m=0,+1\\ a}} \hat{\boldsymbol{\epsilon}}_\alpha \hbar\omega_m\left(\boldsymbol{a}_{m,a}^\dagger \boldsymbol{a}_{m,a} + 1/2\right)|n\rangle_m = \sum_{m=0,+1} \hbar\omega_m(n_m + 1/2) \quad (67)$$

$$\boldsymbol{p}^{\boldsymbol{L}}|n\rangle_m = \frac{h}{4\pi}\sum_{\boldsymbol{m}=\boldsymbol{0},-\boldsymbol{1}} \boldsymbol{k}_m[\boldsymbol{a}_m\boldsymbol{a}_m^\dagger + \boldsymbol{a}_m^\dagger\boldsymbol{a}_m]|n\rangle_m = \sum_{m=0,-1} \hbar\boldsymbol{k}_m n_m \quad (68)$$

$$\boldsymbol{p}^{\boldsymbol{R}}|n\rangle_m = \frac{h}{4\pi}\sum_{\boldsymbol{m}=\boldsymbol{0},+\boldsymbol{1}} \boldsymbol{k}_m[\boldsymbol{a}_m\boldsymbol{a}_m^\dagger + \boldsymbol{a}_m^\dagger\boldsymbol{a}_m]|n\rangle_m = \sum_{m=0,+1} \hbar\boldsymbol{k}_m n_m \quad (69)$$

$$L_z^L|n\rangle_m = \frac{\boldsymbol{L}|n\rangle_m \cdot \boldsymbol{k}_z}{|\boldsymbol{k}_z|} = i\hbar\sum_{m=0,-1} m * n_m = -n_{-1} * i\hbar \quad (70)$$

$$L_z^R|n\rangle_m = \frac{\boldsymbol{L}|n\rangle_m \cdot \boldsymbol{k}_z}{|\boldsymbol{k}_z|} = i\hbar\sum_{m=0,+1} m * n_m = n_{+1} * i\hbar \quad (71)$$

$$\mathbf{S}^L|n\rangle_m = \sum_{m=0,-1,\alpha} \hbar\hat{\boldsymbol{k}}_{m,\alpha}(\boldsymbol{n}_{m,L,\alpha} - \boldsymbol{n}_{m,R,\alpha}) \quad (72)$$

$$\mathbf{S}^R|n\rangle_m = \sum_{m=0,+1,\alpha} \hbar\hat{\boldsymbol{k}}_{m,\alpha}(\boldsymbol{n}_{m,L,\alpha} - \boldsymbol{n}_{m,R,\alpha}) \quad (73)$$

Similarly, when the nanowire is excited by right circularly polarization (RCP) light, the electric field components of RCP ($E_z = E_0, E_x = E_0 e^{i*\frac{\pi}{2}}$ ) illuminate on the nanowire, resulting in an analogous form of the vector potential.

$$\boldsymbol{A}^{RCP}(\boldsymbol{r},t)$$

$$= \boldsymbol{A}_0 e^{i(k_{z,0}z-\omega t)} + \boldsymbol{A}_{+1}\cos(\emptyset)\, e^{i(k_{z,+1}z-\omega t)} e^{i*\frac{\pi}{2}} + \boldsymbol{A}_{+1}\cos\left(\emptyset - \frac{\pi}{2}\right) e^{i\left(k_{z,+1}z-\omega t+\frac{\pi}{2}\right)}$$

$$= \boldsymbol{A}_0 e^{i(k_{z,0}z-\omega t)} + i * \boldsymbol{A}_{+1}\cos(\emptyset)\, e^{i(k_{z,+1}z-\omega t)} + i * \boldsymbol{A}_{+1}\sin(\emptyset)\, e^{i(k_{z,+1}z-\omega t)}$$

$$= \boldsymbol{A}_0 e^{i(k_{\parallel,0}z-\omega t)} + i * \boldsymbol{A}_{+1}(\cos(\emptyset) + \sin(\emptyset)) e^{i(k_{z,+1}z-\omega t)} \quad (74)$$

The term $\cos(\emptyset)+\sin(\emptyset)$ reflects the oblique distribution of the SPPs excited by the circularly polarization light are on the $\emptyset=+\pi/4$ side, which is shown in Fig. 4. For symmetry, the over complete base of m can be used to describe the vector potential.

$$\boldsymbol{A}^{RCP}(\boldsymbol{r},t)=\boldsymbol{A}_0 e^{i(k_{z,0}z-\omega t)}+\frac{1}{2}\left((i+1)\boldsymbol{A}_{+1}e^{i\emptyset}+\boldsymbol{A}_{+1}(\mathrm{i}-1)e^{-i\emptyset}\right)e^{i\left(k_{z,+1}z-\omega t\right)}$$

$$=\boldsymbol{A}_0 e^{i(k_{z,0}z-\omega t)}+\frac{1}{2}((i+1)\boldsymbol{A}_{+1}e^{i\emptyset}-\boldsymbol{A}_{-1}(\mathrm{i}-1)e^{-i\emptyset})e^{i\left(k_{z,+1}z-\omega t\right)} \quad (75)$$

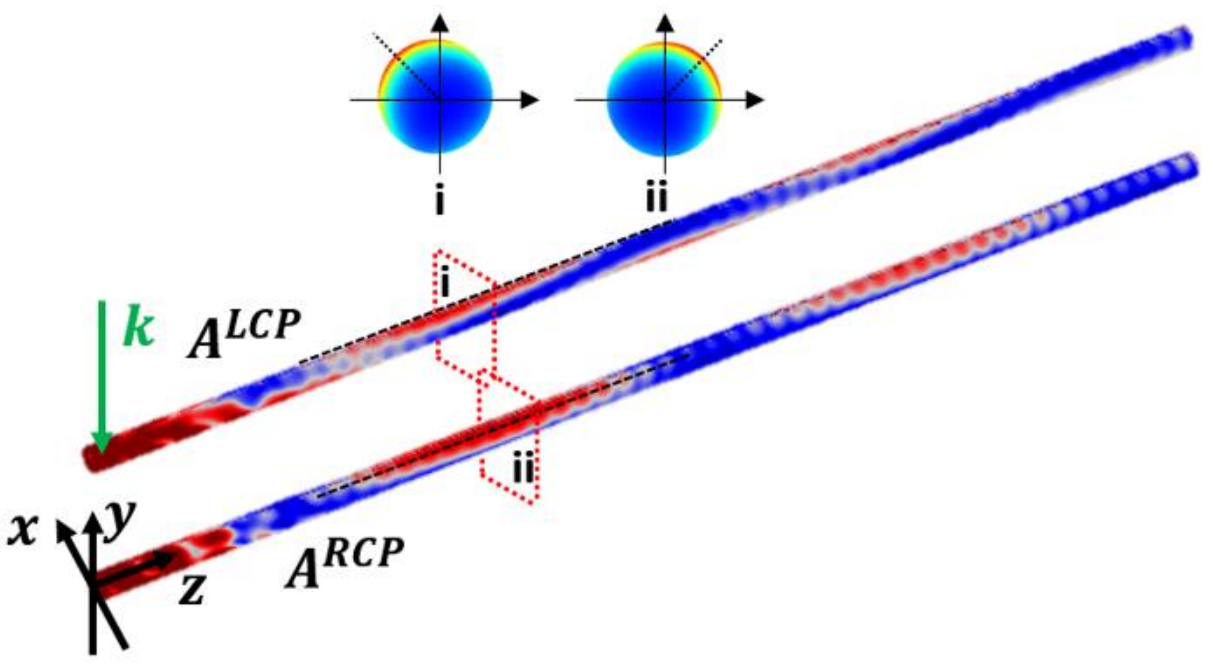

**Fig. 4.** The scheme of electric field distributions on plasmonic cylindrical nanowires excited by focused LCP($\mathrm{A}^{LCP}$) and RCP($\mathrm{A}^{RCP}$) light on left ends. The insets show the cross section at the wires indicated at position i and ii.

From the expression one can see that the pattern should be the superpostion of the two helical modes $m=+1$ and $m=-1$. It is also consistant with the fact that the circularly polarized light can be decomposed with two orthorgnal linearly polarized light for the exciting light.

For left circularly polarized (LCP) light exciation on one end, similarly we have

$$\boldsymbol{A}^{LCP}(\boldsymbol{r},t)=\boldsymbol{A}_0 e^{i(k_{z,0}z-\omega t)}+\boldsymbol{A}_{+1}\cos(\emptyset)\,e^{i\left(k_{z,+1}z-\omega t\right)}e^{-i*\frac{\pi}{2}}+\boldsymbol{A}_{+1}\cos\left(\emptyset-\frac{\pi}{2}\right)e^{i\left(k_{z,+1}z-\omega t+\frac{\pi}{2}\right)}=\boldsymbol{A}_0 e^{i(k_{z,0}z-\omega t)}-i*\boldsymbol{A}_{+1}(\cos(\emptyset)-\sin(\emptyset))e^{i\left(k_{z,+1}z-\omega t\right)} \quad (76)$$

$$\boldsymbol{A}^{LCP}(\boldsymbol{r},t)$$
$$=\boldsymbol{A}_0 e^{i(k_{z,0}z-\omega t)}+\frac{1}{2}\left((-1+i)\boldsymbol{A}_{+1}e^{i\emptyset}+\boldsymbol{A}_{+1}(-1-\mathrm{i})e^{-i\emptyset}\right)e^{i\left(k_{z,+1}z-\omega t\right)}$$

$$=\boldsymbol{A}_0 e^{i(k_{z,0}z-\omega t)}+\frac{1}{2}((-1+i)\boldsymbol{A}_{+1}e^{i\emptyset}+\boldsymbol{A}_{-1}(1+\mathrm{i})e^{-i\emptyset})e^{i\left(k_{z,+1}z-\omega t\right)} \quad (77)$$

The coefficient shows that the pattern of $\boldsymbol{A}^{LCP}$ is the beat at the $\emptyset=-\pi/4$ sides (Fig. 4). And the results in Fock states can be written as following:

$$\mathrm{H}^{LCP}|n\rangle_m=\sum_{\substack{m=0,\pm1\\ \alpha}}\hat{\boldsymbol{\epsilon}}_\alpha\hbar\omega_m\left(\boldsymbol{a}_{m,\alpha}^{\dagger}\boldsymbol{a}_{m,\alpha}+1/2\right)[(+\tfrac{1}{2})^{|m|}|n\rangle_m]$$

$$=\sum_{m=0,\pm1}(+\tfrac{1}{2})^{|m|}\hbar\omega_m(n_m+1/2)=\sum_{m=0,+1}\hbar\omega_m(n_m+1/2) \quad (78)$$

$$\mathrm{H}^{RCP}|n\rangle_m = \sum_{m=0,+1} \hbar\omega_m(n_m + 1/2) \quad (79)$$

$$\boldsymbol{p}^{LCP}|n\rangle_m = \frac{h}{4\pi}\sum_{m=0,\pm1} \boldsymbol{k}_m\left[\boldsymbol{a}_m\boldsymbol{a}_m^{\dagger} + \boldsymbol{a}_m^{\dagger}\boldsymbol{a}_m\right]\left[\left(+\frac{1}{2}\right)^{|m|}|n\rangle_m\right]$$

$$= \sum_{m=0,\pm1}(+\tfrac{1}{2})^{|m|}\hbar c\boldsymbol{k}_m n_m = \sum_{m=0,+1} \hbar\boldsymbol{k}_m n_m \quad (80)$$

$$\boldsymbol{p}^{RCP}|n\rangle_m = \sum_{m=0,+1} \hbar\boldsymbol{k}_m n_m \quad (81)$$

$$L_z^{LCP}|n\rangle_m = \frac{\boldsymbol{L}|n\rangle_m\cdot\boldsymbol{k}_z}{|\boldsymbol{k}_z|} = ih\sum_{m=0,\pm1} m * \widehat{\boldsymbol{n}}_m\left[\left(+\frac{1}{2}\right)^{|m|}|n\rangle_m\right] = 0 \quad (82)$$

$$L_z^{RCP}|n\rangle_m = \frac{\boldsymbol{L}|n\rangle_m\cdot\boldsymbol{k}_z}{|\boldsymbol{k}_z|} = ih\sum_{m=0,\pm1} m * \widehat{\boldsymbol{n}}_m[(+\tfrac{1}{2})^{|m|}|n\rangle_m] = 0 \quad (83)$$

$$\mathbf{S}^{LCP}|n\rangle_m = \sum_{m=0,\pm1,\alpha} \hbar\widehat{\boldsymbol{k}}_{m,\alpha}\left(\boldsymbol{n}_{m,L,\alpha} - \boldsymbol{n}_{m,R,\alpha}\right)\left[\left(+\frac{1}{2}\right)^{|m|}|n\rangle_m\right] \quad (84)$$

$$\mathbf{S}^{RCP}|n\rangle_m = \sum_{m=0,\pm1,\alpha} \hbar\widehat{\boldsymbol{k}}_{m,\alpha}\left(\boldsymbol{n}_{m,L,\alpha} - \boldsymbol{n}_{m,R,\alpha}\right)[(+\tfrac{1}{2})^{|m|}|n\rangle_m] \quad (85)$$

From the expression one can find out that the orbital angular momentum of SPPs on nanowire is zero under circularly polarized light excitation.

## 5. Discussion

During the above process, the electromagnetic modes of the plasmonic cylindrical wire are deduced and quantized, which is consistent with the quantized charge density wave on wire with hydrodynamic method[41, 50] due to all of the charge responses in optics are included in permittivity of the material. One can see that the spin of SPPs on the wire contains both transverse component, which agrees with the conclusion of SPPs on plane surface[47, 51-53]. The longitudinal components may be from the curvature boundary of the wire[54]. The higher modes also carry orbital angular momentum which is very similar with the vortex wave. Along the wire, the orbital angular momentum quantum number can be $0, \pm1, \pm2, \ldots$. Experimentally, it can be verified by putting the nanowire in homogeneous medium and the helix can be shown with quantum dots[16]. The momentum can be observed by using leakage Fourier transform microscope[55]. The spin-based effects also have an unprecedented potential to control light and its polarization, thereby promoting the research of optical chirality. The spin angular momentum and the orbital angular momentum will have coupling[47] and cause some other effects on the nanowire waveguide. In the experiment, the wires are sure with limited length. If the wire is not very long, the SPPs will be reflected by the other end and form standing wave, which is like a Fabry–Pérot cavity[7]. When it interacts with other systems like atoms or quantum dots, some hyperfine phenomenon may be expected[56, 57]. In the above calculations, the dissipation of the nanowire material is ignored. However, if the imaginary part of the permittivity is considered (lossy medium), the commutation relation for canonical coordinate and momentum is not available any more, and the commutation relation for the creation and annihilation operators will have a decay coefficient. When the decay term is absorbed in the operators, all of the quantities will include the decay effect, and the ground state energy is also decaying because of the damping.

## 6. Conclusion

In this work, the electromagnetic modes on plasmonic cylinder waveguide are quantized. The orbital and spin angular momentum is studied and it can be seen that similar to the vortex waves, the plasmon modes on cylinder nanowires carry both orbital angular momentum and spin angular momentum. The results may be helpful for the future quantum information applications.

## Appendix A: The electromagnetic fields of wire

The electric and magnetic fields satisfy the vector Helmholtz equations

$$\nabla^2 \begin{Bmatrix} \boldsymbol{E} \\ \boldsymbol{H} \end{Bmatrix} - \mu\varepsilon \frac{\partial^2}{\partial t^2} \begin{Bmatrix} \boldsymbol{E} \\ \boldsymbol{H} \end{Bmatrix} = 0 \tag{A1}$$

The vector wavefunctions can be generated from the scalar solutions, which are

$$\psi_i = F_{i,m}(k_{i\perp}\rho) e^{im\phi + ik_{\parallel}z} \tag{$A2$}$$

Where $F_1(k_{i\perp}\rho) = H_m(k_{1\perp}\rho)$ and $F_2(k_{i\perp}\rho) = J_m(k_{2\perp}\rho)$, representing outside and inside the cylinder respectively. $k_{i\perp} = \sqrt{k_i^2 - k_{\parallel}^2}$ is the vertical component of the wave vector. Two sets of solutions of the vector Helmholtz equations are given by

$$\mathbf{v}_i = \frac{1}{k_i} \nabla \times (\hat{z}\psi_i) \tag{$A3$}$$

$$\mathbf{w}_i = \frac{1}{k_i} \nabla \times \mathbf{v}_i \tag{$A4$}$$

Where $\hat{z}$ is a unit vector in the $z$ direction.

Then the curl relations of Maxwell's equations require $\boldsymbol{E}$ and $\boldsymbol{H}$ have the following form

$$\boldsymbol{E}(\boldsymbol{r}) = a_i \mathbf{v}_i(\boldsymbol{r}) + b_i \mathbf{w}_i(\boldsymbol{r})$$

$$= \left\{ \begin{matrix} \left[ \frac{im}{k_{j,m}\rho} a_{j,m} F_{j,m}(k_{j,m\perp}\rho) + \frac{ik_{\parallel,m} k_{j,m\perp}}{k_{j,m}^2} b_{j,m} F'_{j,m}(k_{j,m\perp}\rho) \right] \hat{\rho} \\ + \left[ -\frac{k_{j\perp}}{k_j} a_{j,m} F'_{j,m}(k_{j,m\perp}\rho) - \frac{mk_{\parallel}}{k_j^2 \rho} b_{j,m} F_{j,m}(k_{j,m\perp}\rho) \right] \widehat{\emptyset} \\ + \frac{k_{j,m\perp}^2}{k_{j,m}^2} b_{j,m} F_{j,m}(k_{j,m\perp}\rho) \hat{z} \end{matrix} \right\} e^{i(m\emptyset + k_{\parallel,m} z)} \tag{$A5$}$$

$$\boldsymbol{H}_i(\boldsymbol{r}) = -\frac{i}{\omega\mu_0}k_i[a_i\mathbf{v}_i(\boldsymbol{r}) + b_i\mathbf{w}_i(\boldsymbol{r})]$$

$$= \frac{-ik_{j,m}}{\omega\mu_0}\left\{\begin{matrix} \left[\frac{ik_{\parallel,m}k_{j,m\perp}}{k_{j,m}^2}a_{j,m}F'_{j,m}\left(k_{j,m\perp}\rho\right) + \frac{im}{k_{j,m}\rho}b_{j,m}F_{j,m}\left(k_{j,m\perp}\rho\right)\right]\hat{\rho} \\ + \left[-\frac{mk_{\parallel}}{k_j^2\rho}a_{j,m}F_{j,m}\left(k_{j,m\perp}\rho\right) - \frac{k_{j\perp}}{k_j}b_{j,m}F'_{j,m}\left(k_{j,m\perp}\rho\right)\right]\widehat{\emptyset} \\ + \frac{k_{j,m\perp}^2}{k_{j,m}^2}b_{j,m}F_{j,m}\left(k_{j,m\perp}\rho\right)\hat{z} \end{matrix}\right\}e^{i(m\emptyset + k_{\parallel,m}z)} \quad (A6)$$

where $a_i$ and $b_i$ are constant coefficients.

## Appendix B: Angular momentum

This section aims to show the details in the derivation of orbital and spin angular momentum $\boldsymbol{J} = \frac{1}{4\pi c}\int d^3\boldsymbol{r}\,\boldsymbol{r}\times(\boldsymbol{E}\times\boldsymbol{B}) = \frac{1}{4\pi c}\int d^3\boldsymbol{r}[E_\alpha(\boldsymbol{r}\times\boldsymbol{\nabla})A_\alpha + \boldsymbol{E}\times\boldsymbol{A}]$.

a) The derivation of orbital angular momentum $\boldsymbol{L}$

The first term of the right side of $\boldsymbol{J}$ is the orbital part. The orbital angular momentum density is

$$\boldsymbol{l}_\alpha = (\boldsymbol{r}\times\boldsymbol{\nabla})A_\alpha = \begin{pmatrix} -\frac{z}{\rho}\frac{\partial \boldsymbol{A}_\rho}{\partial\emptyset}\,\widehat{\boldsymbol{\rho}} \\ (z\frac{\partial \boldsymbol{A}_\rho}{\partial\rho} - \rho\frac{\partial \boldsymbol{A}_\rho}{\partial z})\widehat{\emptyset} \\ \frac{\partial \boldsymbol{A}_\rho}{\partial\emptyset}\,\hat{\boldsymbol{z}} \end{pmatrix} + \begin{pmatrix} -\frac{z}{\rho}\frac{\partial \boldsymbol{A}_\emptyset}{\partial\emptyset}\,\widehat{\boldsymbol{\rho}} \\ (z\frac{\partial \boldsymbol{A}_\emptyset}{\partial\rho} - \rho\frac{\partial \boldsymbol{A}_\emptyset}{\partial z})\widehat{\emptyset} \\ \frac{\partial \boldsymbol{A}_\emptyset}{\partial\emptyset}\,\hat{\boldsymbol{z}} \end{pmatrix} + \begin{pmatrix} -\frac{z}{\rho}\frac{\partial \boldsymbol{A}_z}{\partial\emptyset}\,\widehat{\boldsymbol{\rho}} \\ (z\frac{\partial \boldsymbol{A}_z}{\partial\rho} - \rho\frac{\partial \boldsymbol{A}_z}{\partial z})\widehat{\emptyset} \\ \frac{\partial \boldsymbol{A}_z}{\partial\emptyset}\,\hat{\boldsymbol{z}} \end{pmatrix}$$

$$= i\sqrt{\frac{2\pi c^2\hbar}{V}}\sum_m \hat{r}_\alpha\cdot\begin{pmatrix} -\frac{z}{\rho}m \\ -\rho k_{z,m} \\ m \end{pmatrix}\left[\boldsymbol{a}_{m\alpha}e^{i\boldsymbol{k}_{\parallel,m}\boldsymbol{r}} - \boldsymbol{a}_{m\alpha}^{\dagger}e^{-i\boldsymbol{k}_{\parallel,m}\boldsymbol{r}}\right] \quad \text{(B1)}$$

Combining with the corresponding electric field component, the orbital angular momentum is

$$\boldsymbol{L} = \frac{1}{4\pi c}\int \boldsymbol{d^3 r} E_\alpha l_\alpha = \frac{i\hbar}{2V}\sum_{m,m',\alpha,\alpha'} \frac{\omega_m}{\sqrt{\omega_m\omega_{m'}}}\hat{\boldsymbol{\epsilon}}_\alpha \cdot \hat{\boldsymbol{\epsilon}}_{\alpha'} \int d^3\boldsymbol{r} \begin{pmatrix} -\frac{z}{R}m \\ -Rk_{z,m} \\ m \end{pmatrix}$$

$$\left[\boldsymbol{a}_{m\alpha} e^{\boldsymbol{i}\boldsymbol{k}_{\parallel,m}\boldsymbol{r}} - \boldsymbol{a}^\dagger_{m\alpha} e^{-\boldsymbol{i}\boldsymbol{k}_{\parallel,m}\boldsymbol{r}}\right]\left[\boldsymbol{a}_{m'\alpha'} e^{\boldsymbol{i}\boldsymbol{k}_{\parallel,m'}\boldsymbol{r}} - \boldsymbol{a}^\dagger_{m'\alpha'} e^{-\boldsymbol{i}\boldsymbol{k}_{\parallel,m'}\boldsymbol{r}}\right] \qquad (B2)$$

We can integrate different component over space:

$$\mathrm{I}_\rho = \int d^3 r \left(-\frac{z}{R} m e^{\boldsymbol{i}\boldsymbol{k}_{\parallel,m'}\boldsymbol{r}} e^{\boldsymbol{i}\boldsymbol{k}_{\parallel,m}\boldsymbol{r}}\right) = i\frac{\partial}{\partial k_z} V\delta_{m,\mathrm{m}'} \qquad (B3)$$

The $\rho$ component of OAM $\boldsymbol{L}$ should be zero.

$$\mathrm{I}_\emptyset = \int d^3 r\left(-Rk_{z,m} e^{\boldsymbol{i}\boldsymbol{k}_{\parallel,m'}\boldsymbol{r}} e^{\boldsymbol{i}\boldsymbol{k}_{\parallel,m}\boldsymbol{r}}\right) = -Rk_{z,m} V\delta_{m,\mathrm{m}'} \qquad (B4)$$

$$\mathrm{I}_z = \int d^3 r\left(m e^{\boldsymbol{i}\boldsymbol{k}_{\parallel,m'}\boldsymbol{r}} e^{\boldsymbol{i}\boldsymbol{k}_{\parallel,m}\boldsymbol{r}}\right) = mV\delta_{m,\mathrm{m}'} \qquad (B5)$$

$$\boldsymbol{L} = \frac{i\hbar}{2}\sum_{m,\alpha} \boldsymbol{l}_{m,\alpha}\left(\boldsymbol{a}_{m,\alpha}\boldsymbol{a}^\dagger_{m,\alpha} + \boldsymbol{a}^\dagger_{m,\alpha}\boldsymbol{a}_{m,\alpha}\right) = \sum_m i\hbar \boldsymbol{l}_m \hat{\boldsymbol{n}}_m \qquad (B6)$$

Where $\boldsymbol{l}_m = \begin{pmatrix} 0 \\ -Rk_{z,m} \\ m \end{pmatrix}$.

b) The derivation of commutation relation of $\boldsymbol{S}$

$$\left[(\boldsymbol{S}_i)(\boldsymbol{S}_j)\right]_{ln} = (\boldsymbol{S}_i)_{lm}(\boldsymbol{S}_j)_{mn} = (-i\hbar)^2 \varepsilon_{ilm}\varepsilon_{jmn} = (-i\hbar)^2\left(\delta_{in}\delta_{lj} - \delta_{ij}\delta_{ln}\right) \qquad (B7)$$

$$\left[(\boldsymbol{S}_j)(\boldsymbol{S}_i)\right]_{ln} = (\boldsymbol{S}_j)_{lm}(\boldsymbol{S}_i)_{mn} = (-i\hbar)^2 \varepsilon_{jlm}\varepsilon_{imn} = (-i\hbar)^2\left(\delta_{jn}\delta_{li} - \delta_{ji}\delta_{ln}\right) \qquad (B8)$$

Subtraction of two formulas

$$\left[(\boldsymbol{S}_i)(\boldsymbol{S}_j) - (\boldsymbol{S}_j)(\boldsymbol{S}_i)\right]_{ln} = (-i\hbar)^2\left(\delta_{in}\delta_{lj} - \delta_{jn}\delta_{li}\right) \qquad (B9)$$

The difference of deltas can be replaced by the product of $\varepsilon$'s

$$\begin{aligned}\left[(\boldsymbol{S}_i)(\boldsymbol{S}_j) - (\boldsymbol{S}_j)(\boldsymbol{S}_i)\right]_{ln} &= (-i\hbar)^2(-\varepsilon_{ijk})\varepsilon_{lnk} \\ &= (i\hbar\varepsilon_{ijk})[-i\hbar\varepsilon_{kln}] = i\hbar\varepsilon_{ijk}(\boldsymbol{S}_k)_{ln} \qquad (B10)\end{aligned}$$

Then the operator $S_i$ do satisfy the commutation relation:

$$[(\boldsymbol{S}_i),(\boldsymbol{S}_j)] = i\hbar\varepsilon_{ijk}(\boldsymbol{S}_k) \qquad (B11).$$

**Funding**

National Natural Science Foundation of China (NSFC) (11704058, 11974069); the National Special Support Program for High-level Personnel Recruitment (W03020231), LiaoNing Revitalization Talents Program (XLYC1902113), Program for Liaoning Innovation Team in University (LT2016011), Science and Technique Foundation of Dalian (Grant No. 2017RD12), Fundamental Research Funds for the Central Universities (DUT19RC(3)007).

**Acknowledgement**

The authors thank Prof. Weijie Fu and Prof. Jiasen Jin's helpful discussion.

**Disclosures**

The authors declare no conflicts of interest.